\documentclass[graybox]{svmult}
\usepackage{booktabs}
\usepackage{array}
\usepackage{amsmath}
\usepackage{amsfonts}
\usepackage{amssymb}
\usepackage{mathptmx}
\usepackage{helvet}
\usepackage{courier}
\usepackage{type1cm}
\usepackage{makeidx}
\usepackage{graphicx}
\usepackage{multicol}
\usepackage{xcolor}
\usepackage[bottom]{footmisc}

\makeindex

\newcommand{\yvet}{\vec{y}}

\newcommand{\Cvet}{\vec{C}}

\newcommand{\Ivet}{\vec{I}}

\newcommand{\Kvet}{\vec{K}}

\newcommand{\Mvet}{\vec{M}}

\newcommand{\Wvet}{\vec{W}}

\newcommand{\Yvet}{\vec{Y}}

\newcommand{\Unovet}{\vec{1}}
\newcommand{\Zerovet}{\vec{0}}

\newcommand{\epsvet}{\boldsymbol{\varepsilon}}

\newcommand{\Omegavet}{\boldsymbol{\Omega}}

\DeclareMathOperator*{\argmin}{arg\,min}

\begin{document}

\title*{Forecasting Italian daily electricity generation disaggregated by geographical zones and energy sources using coherent forecast combination}
\titlerunning{Forecasting Italian daily electricity generation using coherent forecast combination} 

\author{Daniele Girolimetto and Tommaso Di Fonzo}

\institute{Department of Statistical Sciences (University of Padova) Via Cesare Battisti, 241, 35121 Padova, \email{\texttt{daniele.girolimetto@unipd.it}}}

\maketitle

\abstract{A novel approach is applied for improving forecast accuracy and achieving coherence in forecasting the Italian daily energy generation time series. In hierarchical frameworks such as national energy generation disaggregated by geographical zones and energy sources, independently generated base forecasts often result in inconsistencies across the constraints. We deal with this issue through a coherent  balanced multi-task forecast combination approach, which combines unbiased forecasts from multiple experts while ensuring coherence. Applied to the daily Italian electricity generation data, our method shows superior accuracy compared to single-task base and combined forecasts, and a state-of-the-art single-expert reconciliation technique, demonstrating to be an effective approach to forecasting linearly constrained multiple time series.}

\vspace*{-1.5em}
\section{Introduction}
\vspace*{-1em}
\label{sec:intro}

Selecting an effective forecasting method among many options presents a complex and resource-intensive challenge, as inherent uncertainties often prevent any method from achieving the best accuracy across all scenarios. This complexity becomes even more significant when dealing with multiple interrelated variables that must satisfy linear constraints, as in National Accounts \cite{Girolimetto2024} and in hierarchical time series like electricity demand segmented by regions and sources \cite{Athanasopoulos2024}. Such settings require appropriate forecasting approaches ensuring coherence across interrelated variables. Traditional univariate forecast combination methods \cite{Bates1969, Clemen1989} focus on individual time series. In contrast, global forecast combination methods leverage shared information across multiple series to improve accuracy \cite{Thompson2024}. Similarly, forecast reconciliation approaches search for coherence by adjusting forecasts to satisfy some \textit{a priori} constraints \cite{Wickramasuriya2019}, with significant empirical advantages in various fields \cite{Athanasopoulos2024}. 

In this paper, we exploit a recently developed linear framework \cite{DGTDF2025}, that joins single-variable forecast combination \cite{Bates1969} and the least squares reconciliation procedure for a linearly constrained multiple time series \cite{Stone1942-fa} to produce optimal (in the least squares sense) coherent forecasts. The remainder of the paper is structured as follows. Sect.~\ref{sec:comb} succinctly presents the optimal coherent forecast combination procedure for incoherent base forecasts of a linearly constrained multiple time series, of which hierarchical time series are a special case. Sect.~\ref{sec:energy} presents the main findings of a forecasting experiment on 76 daily time series of the Italian electricity generation disaggregated by geographical zones and energy sources.

\vspace*{-2.5em}
\section{Optimal coherent multi-task forecast combination}\label{sec:comb}
\vspace*{-1em}

Let $\yvet = \left[y_1 \; \ldots \; y_i \; \ldots \; y_n\right]^\top \in \mathbb{R}^n$ represent the target forecast vector for a linearly constrained time series $\Yvet_t$, such that $\Cvet \Yvet_t = \Zerovet_{(n_u \times 1)}$, where $\Cvet \in \mathbb{R}^{n_u \times n}$ is a predefined matrix describing the (homogeneous) linear constraints valid for the single component series of $\Yvet_t$, i.e., $\Cvet\yvet = \Zerovet_{(n_u \times 1)}$. We assume that $p \geq 2$ experts independently produced a base forecast for every one of the $n$ individual components of $\yvet$, yielding $m = np$ base forecasts overall\footnote{A complete discussion of the optimal coherent combination approach, including the extension to the general case where the experts may produce base forecasts only for a subset of the totality of variables, can be found in \cite{DGTDF2025}.}.

Denote $\widehat{y}_i^j$ the unbiased base forecast of the $i$-th individual variable of $\yvet$ produced by the $j$-th expert, and let $\varepsilon_i^j = \widehat{y}_i^j - y_i$ the zero-mean base forecast error. Let $\widehat{\yvet}^j \in \mathbb{R}^{n_j}$, $j=1,\ldots,p$, be the vector of base forecasts produced by the $j$-th expert, with error given by $\epsvet^j = \widehat{\yvet}^j - \yvet$, $\widehat{\yvet} = \text{vec}\begin{bmatrix}
	\widehat{\yvet}^1 & \dots & \widehat{\yvet}^j & \dots & \widehat{\yvet}^{p}
\end{bmatrix}  \in \mathbb{R}^m $, and $\epsvet = \text{vec}\begin{bmatrix}
\epsvet^1 & \dots & \epsvet^j & \dots & \epsvet^{p}
\end{bmatrix} \in \mathbb{R}^m $. The linear relationship linking all the available base forecasts $\widehat{\yvet}$ and the unknown target forecast vector $\yvet$ can be expressed through the multiple regression model $\widehat{\yvet} = \Kvet\yvet + \epsvet$, where $\Kvet = \left(\Unovet_p \otimes \Ivet_n\right)$, $E(\epsvet) = \Zerovet_{(np \times 1)}$ and $\Wvet =E(\epsvet\epsvet^\top) \in \mathbb{R}^{m \times m}$ is a p.d. matrix.

We are looking for a coherent forecast vector $\widetilde{\yvet}^c$, i.e., $\Cvet\widetilde{\yvet}^c = \Zerovet_{(n_u \times 1)}$, which exploits all the available base forecasts and improves their accuracy. Forecast combination and reconciliation can be simultaneously dealt with through an optimization-based technique that combines multiple individual experts' base forecasts $\widehat{\yvet}^j$:
\vspace*{-0.5em}
\begin{equation}
	\label{eq:lcquadprog_be}
	\widetilde{\yvet}^c = \argmin_{\yvet}\big(\,\widehat{\yvet} - \Kvet\yvet\big)^\top\Wvet^{-1}\big(\,\widehat{\yvet} - \Kvet\yvet\big) \qquad \text{s.t. } \quad\Cvet\yvet=\Zerovet_{(n_u \times 1)} .
	\vspace*{-0.5em}
\end{equation}
By extending
the well known least squares adjustment procedure for a single vector of preliminary incoherent estimates (i.e., base forecasts) \cite{Stone1942-fa} to the case of $p \ge 2$ vectors, the coherent combined forecast vector $\widetilde{\yvet}^c$ can be expressed as $\widetilde{\yvet}^c = \Mvet \Omegavet^\top\widehat{\yvet}$, where $\Mvet = \left[\Ivet_n - \Wvet_c\Cvet^\top\left(\Cvet\Wvet_c\Cvet^\top\right)^{-1}\Cvet\right]$, $\Omegavet = \Wvet^{-1}\Kvet\Wvet_c$, and $\Wvet_c = \left(\Kvet^\top\Wvet^{-1}\Kvet\right)^{-1}$ (see \cite{DGTDF2025} for details). It is worth noting that this result is obtained under the commonly assumed hypotheses that the base forecasts are unbiased, and that the joint dispersion matrix of the $p$ forecast experts is known.

\vspace*{-2em}
\section{Forecasting Italian daily electricity generation}\label{sec:energy}
\vspace*{-1em}

In this section, we consider a forecasting experiment for $76$ Italian daily electricity generation time series from $01/01/2023$ to $31/12/2023$, obtained from the ENTSO-E Transparency Platform. The Italian electricity market is a grouped time series with two hierarchies, both sharing the top and bottom levels. In the first hierarchy, generation is aggregated into 7 different sources, classified into two categories: renewable (Biomass, Hydro, Solar, Wind, Geothermal) and non-renewable (Fossil, Waste). The second hierarchy organizes the data geographically in 7 zones according to the Terna electricity market map in compliance with the EU CACM regulation.

In order to assess the forecast accuracy of the various approaches, we perform a rolling forecast experiment with an expanding window ($01/01/2023-21/5/2023$ first training set). One- to seven-step-ahead forecasts were generated leading to $Q_1 = 225$ one-, ..., and $Q_7 = 119$ seven-step-ahead daily forecasts for evaluation. Each series was independently modeled using three different approaches: stlf \cite{Cleveland1990-ux}, arima \cite{Box1976-ud}, tbats \cite{De-Livera2011-hx}.
Moreover, we consider the following eleven forecasting procedures:
\begin{enumerate}
	\item \textit{Single best model and its reconciled counterpart}: two approaches. For the best base forecast expert (tbats), the reconciled forecasts are obtained through the shrunk MinT approach by \cite{Wickramasuriya2019} (tbats$_{\text{shr}}$).
	\item \textit{Single-task combination}: three approaches. Equal weights (ew) and the optimal weighting schemes proposed by \cite{Bates1969} and \cite{Newbold1974}, called respectively ow$_{\text{var}}$ and ow$_{\text{cov}}$.
	\item \textit{Coherent combination}: six approaches.
	A sequential reconciliation-then-equal-weight-combination (src),
	three sequential combination-then-reconciliation (scr$_{\text{ew}}$, scr$_{\text{var}}$, scr$_{\text{cov}}$), and finally two optimal coherent combination approach, occ$_{\text{wlsv}}$ and occ$_{\text{be}}$, with diagonal and block-diagonal shrunk in-sample covariance matrix, respectively (see \cite{DGTDF2025} for details).
\end{enumerate}

\noindent The forecast accuracy is evaluated in Table \ref{tab:res} using the Average Relative Mean Absolute ($AR\text{-}MAE$) and Squared ($AR\text{-}MSE$) Error \cite{Fleming1986, Davydenko2013} computed as, respectively, $AR\text{-}MAE_h^{app} = \left(\scalebox{.8}{$\displaystyle\prod_{i = 1}^{76}$} \frac{MAE_{h,i}^{app}}{MAE_{h,i}^{\text{ew}}}\right)^{\frac{1}{76}}$ and $AR\text{-}MSE_h^{app} = \left(\scalebox{.8}{$\displaystyle\prod_{i = 1}^{76}$} \frac{MSE_{h,i}^{app}}{MSE_{h,i}^{\text{ew}}}\right)^{\frac{1}{76}}$,
with $MSE_{h,i}^{app} =\frac{1}{Q_h}\scalebox{.8}{$\displaystyle\sum_{q = 1}^{Q_h}$} \big(y_{i,h,q}-\overline{y}_{i,h,q}^{\,app}\big)^2$ and $MAE_{h,i}^{app} = \frac{1}{Q_h}\scalebox{.8}{$\displaystyle\sum_{q = 1}^{Q_h}$} \big|y_{i,h,q}-\overline{y}_{i,h,q}^{\,app}\big|$, where $h = 1,...,7$ is the forecast horizon, $i=1,...,76$ denotes the variable, $Q_h$ is the test set dimension, $app$ is the approach used, $y_{i,h,q}$ is the observed value and $\overline{y}_{i,h,q}^{\,app}$ is the forecast value using the $app$ approach (coherent or incoherent).  The equal weights (ew) single-task combination is used as benchmark.

Looking at Table~\ref{tab:res}, single-task combination and single-expert reconciliation strategies consistently improve forecasting accuracy compared to the three base models. However, tbats$_{\text{shr}}$ tends to perform less effectively when compared to the more advanced single-task combination approaches. In contrast, coherent combination approaches consistently outperform all the other forecasting strategies. Furthermore, these coherently combined forecasts satisfy all the necessary constraints, which is a critical advantage in this, as in most applied contexts. 

\vspace*{-1.25em}
\begin{acknowledgement}
	The authors acknowledge financial support from project PRIN 2022: PRICE -- \textquotedblleft A New Paradigm for High-Frequency Finance\textquotedblright -- 2022C799SX. 
	\vspace*{-2em}
\end{acknowledgement}

\begin{table}[t]
	\caption{$AR\text{-}MAE$ and $AR\text{-}MSE$ with equal weights combination approach (ew) as benchmark. Bold entries identify the best performing
		approaches, italic entries identify the second best.} 
	\vspace*{-1em}
	\centering
	\setlength{\tabcolsep}{2.5pt}
	\label{tab:res}
\begin{tabular}[t]{l|cccccc|cccccc}
	\toprule
	\multicolumn{1}{c|}{\textbf{}} & \multicolumn{6}{c|}{\textbf{\textbf{AR\text{-}MAE}}} & \multicolumn{6}{c}{\textbf{\textbf{AR\text{-}MSE}}} \\[0.25em]
	\multicolumn{1}{l|}{\textbf{Approach}} & 1 & 2 & 3 & 5 & 7 & 1:7 & 1 & 2 & 3 & 5 & 7 & 1:7\\
	\midrule
	\addlinespace[0.25em]
	\multicolumn{13}{l}{\textit{Base (incoherent forecasts) and single model reconciliation}}\\
	tbats & \textcolor{red}{1.038} & \textcolor{red}{1.042} & \textcolor{red}{1.041} & \textcolor{red}{1.034} & \textcolor{red}{1.031} & \textcolor{red}{1.036} & \textcolor{red}{1.072} & \textcolor{red}{1.078} & \textcolor{red}{1.073} & \textcolor{red}{1.056} & \textcolor{red}{1.041} & \textcolor{red}{1.059}\\
	tbats$_{\text{shr}}$ & \textcolor{red}{1.008} & \textcolor{red}{1.009} & \textcolor{red}{1.018} & \textcolor{red}{1.016} & \textcolor{red}{1.017} & \textcolor{red}{1.014} & \textcolor{red}{1.004} & \textcolor{red}{1.005} & \textcolor{red}{1.029} & \textcolor{red}{1.022} & \textcolor{red}{1.018} & \textcolor{red}{1.018}\\
	\addlinespace[0.25em]
	\multicolumn{13}{l}{\textit{Combination (incoherent forecasts)}}\\
	ew & 1.000 & 1.000 & 1.000 & 1.000 & 1.000 & 1.000 & 1.000 & 1.000 & 1.000 & 1.000 & 1.000 & 1.000\\
	ow$_{\text{var}}$ & 0.988 & 0.987 & 0.991 & 0.994 & 0.996 & 0.992 & 0.978 & 0.978 & 0.986 & 0.991 & 0.997 & 0.989\\
	ow$_{\text{cov}}$ & \textcolor{red}{1.021} & \textcolor{red}{1.015} & \textcolor{red}{1.021} & \textcolor{red}{1.017} & \textcolor{red}{1.006} & \textcolor{red}{1.015} & \textcolor{red}{1.022} & \textcolor{red}{1.020} & \textcolor{red}{1.036} & \textcolor{red}{1.039} & \textcolor{red}{1.039} & \textcolor{red}{1.035}\\
	\addlinespace[0.25em]
	\multicolumn{13}{l}{\textit{Coherent combination}}\\
	src & 0.980 & 0.982 & 0.987 & 0.992 & 0.995 & 0.989 & 0.959 & 0.962 & \em{0.977} & \em{0.985} & 0.994 & \em{0.981}\\
	scr$_{\text{ew}}$ & 0.983 & 0.985 & 0.992 & 0.996 & 0.998 & 0.993 & 0.965 & 0.969 & 0.986 & 0.991 & 0.997 & 0.987\\
	scr$_{\text{var}}$ & \em{0.974} & \em{0.978} & \em{0.986} & 0.993 & 0.995 & \em{0.988} & \em{0.950} & \em{0.958} & 0.977 & 0.989 & 0.998 & 0.982\\
	scr$_{\text{cov}}$ & \textcolor{red}{1.008} & \textcolor{red}{1.009} & \textcolor{red}{1.014} & \textcolor{red}{1.015} & \textcolor{red}{1.005} & \textcolor{red}{1.011} & 0.994 & \textcolor{red}{1.003} & \textcolor{red}{1.024} & \textcolor{red}{1.035} & \textcolor{red}{1.037} & \textcolor{red}{1.026}\\
	occ$_{\text{wls}}$ & 0.984 & 0.982 & 0.986 & \em{0.991} & \em{0.992} & 0.988 & 0.970 & 0.970 & 0.978 & 0.985 & \textbf{0.992} & 0.983\\
	occ$_{\text{be}}$ & \textbf{0.970} & \textbf{0.973} & \textbf{0.979} & \textbf{0.988} & \textbf{0.992} & \textbf{0.983} & \textbf{0.940} & \textbf{0.946} & \textbf{0.966} & \textbf{0.981} & \em{0.993} & \textbf{0.974}\\
	\bottomrule
\end{tabular}
\vspace*{-1em}
\end{table}
\vspace*{-4.5em}

\bibliographystyle{abbrv}
\bibliography{biblio.bib}
\end{document}